\documentclass[twocolumn,a4paper,superscriptaddress,prb]{revtex4}

\usepackage{amsmath}
\usepackage{graphicx}
\usepackage[english]{babel}

\begin{document}

\title{\textit{Ab initio} study of bilateral doping \\ within the MoS$_2$-NbS$_2$ system}

\author{Viktoria V. Ivanovskaya}
\email{ivanovskaya@lps.u-psud.fr}
\affiliation{Laboratoire de Physique des Solides, Univ. Paris-Sud, CNRS-UMR 8502, 91405, Orsay, France}
\affiliation{Institute of Solid State Chemistry, Ural division of Russian Academy of Science, 620041, Ekaterinburg, Russia}

\author{Alberto Zobelli}
\affiliation{Laboratoire de Physique des Solides, Univ. Paris-Sud, CNRS-UMR 8502, 91405, Orsay, France}

\affiliation{Laboratoire des Solides Irradi\'es UMR 7642, CNRS-CEA/DSM, \'Ecole Polytechnique, F-91128 Palaiseau, France}

\author{Alexandre Gloter}
\affiliation{Laboratoire de Physique des Solides, Univ. Paris-Sud, CNRS-UMR 8502, 91405, Orsay, France}

\author{Nathalie Brun}
\affiliation{Laboratoire de Physique des Solides, Univ. Paris-Sud, CNRS-UMR 8502, 91405, Orsay, France}

\author{Virginie Serin}
\affiliation{CEMES-CNRS, 29 rue Jeanne Marvig, BP 94347, Toulouse Cedex 4, France}

\author{Christian Colliex}
\affiliation{Laboratoire de Physique des Solides, Univ. Paris-Sud, CNRS-UMR 8502, 91405, Orsay, France}

\begin{abstract}

We present a systematic study on the stability and the structural and electronic properties of mixed molybdenum-niobium disulphides. Using density functional theory we investigate  bilateral doping with up to 25 \% of MoS$_2$ (NbS$_2$) by Nb (Mo) atoms, focusing on the precise arrangement of dopants within the host lattices. We find that over the whole range of considered concentrations, Nb doping of MoS$_2$ occurs through a substitutional mechanism. For Mo in NbS$_2$  both interstitial and substitutional doping can co-exist, depending upon the particular synthesis conditions. The analysis of the structural and electronic modifications of the perfect bulk systems due to the doping is presented. We show that substitutional Nb atoms introduce electron holes to the MoS$_2$, leading to a semiconductor-metal transition. On the other hand, the Mo doping of Nb$_2$, does not alter the metallic behavior of the initial system. The results of the present study  are compared with available experimental data on mixed MoS$_2$-NbS$_2$ (bulk and nanoparticles).
\end{abstract}

\maketitle

\section{Introduction}

Layered transition metal dichalcogenides belong to a well defined chemical and structural family characterized by strong covalent intralayer bonding and weak Van der Waals interactions between adjacent layers.\cite{Levy,Yoffe-73}
The layer-type structure and the existence of a wide structural gap between planes of metal dichalcogenides offer the possibility of an easy intercalation with a variety of atoms and molecules. \cite{Levy, Starnberg-00}
Despite their structural similarity, metal dichalcogenides present a wide variety of electronic behaviors going from insulators (HfS$_2$) to semiconductors (MoS$_2$), semimetals (TcS$_2$) and real metals (NbS$_2$).\cite{wilson-69} However, it has been shown that the specific electronic properties of metal dichalcogenides can be easily modified by low doping levels.\cite {Kalikhman-82,Me-2}

During the past decades, a number of experimental and theoretical studies have been dedicated to these materials, mainly due to the wide prospects of their potential uses.
The interest in this class of materials has been recently renewed through the synthesis of number of nanosized forms, such as fullerenes and nanotubes, based on \textit{d}-metal disulphides. Their exceptional mechanical properties and a large surface to volume ratio makes them good candidates for new tribological and catalytic applications.\cite{Topsoe-07, Helveg-MoS2Platelets, Tenne-Rev} Besides the pure metal disulphide fullerenes and nanotubes, various ``mixed'' materials and composites have been recently synthesized, such as Nb$_x$W$_{1-x}$S$_{2}$\cite{Zhu-01,Zhu-another-01} and Ti-doped MoS$_2$ nanotubes.\cite{Hsu-01} The synthesis of such mixed systems might be considered as an extension of solid solutions well known from the bulk form into the field of nanomaterials, and represents a promising way of creating functionalized nanoparticles.

In the class of mixed metal dicalcogenites, the MoS$_2$-NbS$_2$ system presents new promising technological potential. Molybdenum disulphide (MoS$_2$) is widely used as catalyst in fuel desulphurization processes and tribological applications \cite{ErtlB,TopsoeB,MoS2-MECH,MoS2-MECH2} and niobium disulphide (NbS$_2$) has attracted attention due to its optical, magnetic and superconductive properties.\cite{wilson-69,Levy,Friend-87,NbS2-InorgChem} The synthesis of mixed  MoS$_2$-NbS$_2$ with no miscibility gap has been reported\cite{Binnewies-05} and for low level Nb-doping of MoS$_2$ particles, a semiconductor-metal transition has been observed.\cite{Kalikhman-82}
The catalytic properties of Mo$_x$Nb$_{1-x}$S$_2$ have been studied in hydrogenation and hydrodesulphurization reactions. In contrast with pure MoS$_2$, the solid solution Mo$_x$Nb$_{1-x}$S$_2$ ($x = 0.4$) has benn shown to be insensitive to H$_2$S partial pressure, thus suggesting good capabilities for the conversion of high sulphur-loaded gas oils.\cite{Catal}
Recently, inorganic fullerene-like Mo$_{1-x}$Nb$_{x}$S$_{2}$ nanoparticles have been synthesized and characterized\cite{Deepack-07} and we expect that the catalytic activity of the mixed phase might be enhanced in the nanoparticles due to a high surface to volume ratio.

In the context of this revived interest in mixed dichalcogenide based systems, the location and local organization of dopant atoms within the original layered lattice play a role of particular importance. However, current analytical techniques can not easily detect the exact arrangement of atoms  in mixed systems, and so neither doping by inter-layer intercalation nor intra-plane substitution can be excluded.\cite{Hsu-01, Nath-02, Deepack-07} Complementary structural information can be provided by simulations using \emph{ab-initio} techniques.

In the this paper we present a density functional theory study of the mixed  MoS$_2$-NbS$_2$ system. We consider bilateral doping of up to 25 \% starting from pure molybdenum disulphide and niobium disulphide bulk systems. Bilateral implies the doping of each metal disulphide by the metal from the counterpart disulphide (i.e. we dope MoS$_2$ with Nb atoms and vice versa, NbS$_2$ with Mo). The paper focus on the arrangement of dopant atoms within the mixed structures and on the electronic modifications induced in the perfect bulk systems due to doping.

\section{Details of the calculations}

\subsection{Computational method}

We performed structural optimizations and electronic structure calculations within the framework of the density function theory in the local density approximation (DFT-LDA) as implemented in the AIMPRO code.\cite{Aimpro, Aimp-2} Molybdenum, niobium and sulphur pseudopotentials are generated using the Hartwingster-Goedecker-Hutter scheme. \cite{Hgh-98}
The basis sets employed consisted of s, p, and d gaussian orbital functions centered at the atomic sites. For Nb and Mo atoms we have used a set of 40 gaussians multiplied by spherical harmonics up to a maximum angular momentum $l_{\text{max}}=2$, for S atoms a set of 20 gaussians with a maximum angular momentum $l_{\text{max}}=1$.

Calculations for bulk MoS$_2$ and NbS$_2$ have been conducted using a 10x10x10 Monkhorst-Pack k-points set to sample the  Brillouin zone.
Defective structures have been described using a 4x4x1 supercell.
2x2x4 Monkhorst-Pack k-points set have been proved to be sufficient for accurate Brillouin zone integration in this large supercell. The atomic positions for all structures were optimized simultaneously with the cell parameters.

\subsection{Structural models \label{structModel}}

\begin{figure}
 \includegraphics[width=\linewidth]{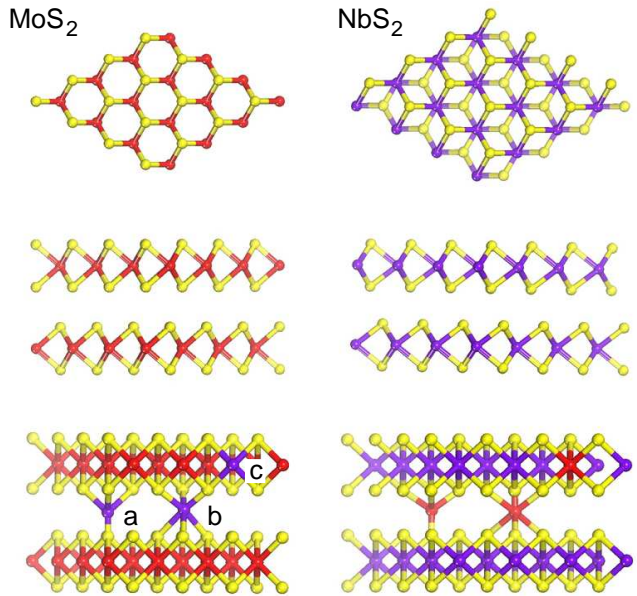}
  \caption{(Color online) Top and side views of the 2H-MoS$_2$ and  2H-NbS$_2$ crystal structures. Different models of doping presented on the bottom pictures: interstitial atoms in  tetrahedral (a) and  octahedral (b) position;  substitutional atom (c).}
  \label{MoS2structure}
\end{figure}

MoS$_2$ crystallizes in several polymorphs, among which the most stable is the hexagonal 2H-MoS$_2$, belonging to the P6$_3$/mmc symmetry point group. The lattice structure consists of hexagonal planes of Mo atoms lying  between two hexagonal planes of S atoms, forming a S-Mo-S layer. Each Mo atom is covalently bonded with its six first neighbor sulphur atoms, each S is bonded to three molybdenum atoms. The unit cell contains two alternating S-Mo-S layers with an ABA BAB stacking along the c axis, see Fig. \ref{MoS2structure}. In bulk 2H-NbS$_2$ the local structure of the layers  is similar to that of 2H-MoS$_2$, with the Nb atoms 6 fold coordinated. The only difference between the structures is that the NbS$_2$ layers are shifted and Nb atoms lie above each other, thus the stacking order within the bulk structure is ACA BCB, see Fig \ref{MoS2structure}.

The experimental lattice parameters for 2H-MoS$_2$  and 2H-NbS$_2$ metal disulphides  are  a = 3.16 \AA, c = 12.29 \AA, and a = 3.31 \AA, c = 11.89 \AA, respectively \cite{Levy,wilson-69}. Molybdenum and niobium oxidation states in the corresponding sulfate bulk systems are identical and the covalent radii of Mo (130 pm) and Nb (132 pm) are similar, implying that bilateral substitution might be feasible.
Furthermore, the interlayer spacing for both metal dichalcogenides is wide enough to place the interstitial atom in between the layers.
Hence, in the present work we will analyze both, substitutional and interstitial doping modes for accommodating the metal atoms within the layered disulphide crystals.

For the interstitial doping we have to distinguish two cases: the dopant atom occupies  the tetrahedral site situated  between one sulphur triangle and one sulphur atom  in the adjacent plane (position a, Fig. \ref{MoS2structure}) or the octahedral site between two S-triangles (position b, Fig. \ref{MoS2structure}). In the first case only four metal-sulphur bonds are formed, whereas in the second case six. In the substitutional configuration one metal atom is removed from the metal disulphide (MS$_2$) layer and the vacant site is occupied by a dopant atom, which thus forms six metal-sulphur bonds (position c, Fig. \ref{MoS2structure}).

\section{Results and Discussion}

\subsection{Bulk systems}

The optimized  bulk cell parameters are in excellent agreement with the experimental values: a = 3.15 \AA, c = 12.29 \AA \- for MoS$_2$ and a = 3.32 \AA, c = 11.92 \AA \- for NbS$_2$. The Mo-S and Nb-S bond lengths in the pure bulk systems are found to be 2.42 \AA \- and 2.50 \AA, which is close to the reported experimental values respectively of 2.41 \AA \- and 2.47 \AA. \cite{wilson-69}

\begin{figure}[btp]
\includegraphics[bb=50 76 500 378 ,angle=0,width=\columnwidth]{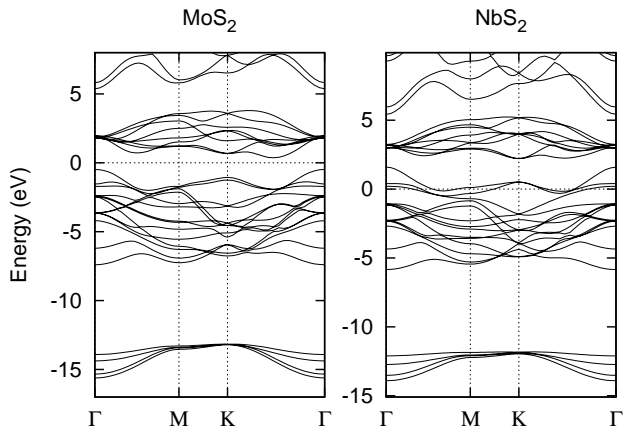}
\caption{DFT-LDA electronic band structure for pure MoS$_2$ and NbS$_2$. The origin of the energies is the Fermi level.}
\label{pureElstruct}
\end{figure}

The electronic structure of bulk metal dichalcogenides  have been extensively studied both experimentally and theoretically. \cite{wilson-69,Friend-87, Raybaud-97}
In Fig.\ref{pureElstruct} we present the electronic band structures for MoS$_2$ and NbS$_2$ bulk species. The deepest band in the plot corresponds to the S 3s states and is separated by a wide gap from a broad valence band.  The bottom of the valence band corresponds to the hybrid 3$p$-S and 4$d$-Mo(Nb) states, and the upper part to the  4$d_{z^2}$ Mo(Nb) states. As one can notice, the overall structure and dispersion of the bands for the MoS$_2$ and NbS$_2$ bulks are rather similar.
The main difference between the two systems comes from the different number of valence electrons between the Mo and Nb atoms. The number of valence electrons in MoS$_2$ is enough to fill completely the valence band and thus MoS$_2$ is a semiconductor with an LDA indirect gap of 1.17  eV (experimentally measured\cite{Kam-81} as 1.23 eV) and a direct gap at the $\Gamma$ point of 2.02 eV (experimentally\cite{Kam-81} 1.74 eV). However, NbS$_2$ has one  electron less per metal atom so that the top of its valence band is half-filled. Consequently, NbS$_2$ has a metallic behavior. Our results for bulk systems agree well with the previously reported calculations and available experimental data. \cite{DeGroot-87,Kobayashi-95,Boker-01,Kim-05, Fang-95}

\subsection{Low concentration doping: single dopant atoms\label{secLowcon}}

In order to  access the most energetically favorable way for bilateral doping within the MoS$_2$ -NbS$_2$ system, we have calculated the formation energies of the ``mixed'' structures. For both types of doping, the formation energy can be estimated from the general equation

\begin{equation}
E_{\text{form}}=E_{\text{mixed}}-[M\cdot\mu_{\text{MoS}_2}+N\cdot\mu_{\text{NbS}_2}+X\cdot\mu_{\text{D}}]
\label{EformInter}
\end{equation}

where

$$\left\{\begin{array}{ll}
M,N \neq 0 \text{ and } X=0 & \text{for substitutional doping};\\
M\text{ or }N=0\text{ and }X \neq 0 & \text{for interstitial doping}.
\end{array}\right.$$

In Eq.\ref{EformInter} E$_{\text{mixed}}$ is the total energy of the mixed system; M and N are the numbers of MoS$_2$ and NbS$_2$ units in the mixed systems; $\mu_{\text{MoS}_2}$ and  $\mu_{\text{NbS}_2}$ are respectively the chemical potentials for MoS$_2$ and NbS$_2$ using as reference the corresponding bulk systems; X is the number of dopant atoms inside the cell and $\mu_D$ is the chemical potential of the dopant atom.

As a general rule, in the synthesis process the chemical potential  for an element as a dopant should be lower than that in its bulk form, otherwise this element would form the energetically more stable bulk phase rather than the mixed system. Consequently, in our study  we imposed as boundaries the chemical potentials of dopants derived from the corresponding pure bulk metals $\mu_{\text{D}}=\mu_{\text{bulk}}$ or metal disulphides $\mu_D = \mu_{\text{MS}_2}-2 \cdot \mu_{\text{S}}$, where $\mu_S$ is sulphur chemical potential as derived from its bulk phase. Thus, we obtain the range of formation energies calculated using both values of the dopant's chemical potentials for each dopant configuration.

\begin{table}[bp]
\caption{Formation energies (eV) and average metal-sulphur bond distances (\AA) in single atom doped MoS$_2$ and NbS$_2$.
Formation energies are calculated using the dopant chemical potentials derived from metal disulphide (E$_{form_1}$) or pure metal bulk (E$_{form_2}$) phases.}
\begin{tabular*}{\linewidth}{@{\extracolsep{\fill}}lrrrr}
\hline
\hline
 System & E$_{form_1}$  & E$_{form_2}$  & Mo-S  & Nb-S  \\
\hline
 MoS$_2$&     && 2.42    &    \\
 NbS$_2$&     & &    &  2.50  \\
\hline
 NbMo$_{32}$S$_{64}$ (inter) & 8.08& 3.80 & 2.42 & 2.43    \\
 NbMo$_{31}$S$_{64}$ (subst) &-0.21& -0.86 &  2.42 & 2.45  \\
 \hline
 MoNb$_{32}$S$_{64}$ (inter) &  -3.24 & -6.87 &  2.40  &  2.49   \\
 MoNb$_{31}$S$_{64}$ (subst) & -5.11 & -4.46  &  2.46 &  2.49  \\
\hline
\hline
\end{tabular*}
\label{eformTab}
\end{table}

Formation energies for low level doping (one dopant per unit cell, which corresponds to $\sim 3$\% of doping) and metal-sulphur bond lengths are presented in Table \ref{eformTab}. The inclusion of Nb dopant in interstitial sites within the MoS$_2$ structure is highly energetically unfavorable since the formation energy for the Nb in octahedral sites is of the order of several eV. The Nb  position in tetrahedral sites (position a, see Fig.\ref{MoS2structure}) is not stable and after relaxation the Nb atom moves to a neighboring octahedral site (position b).
In contrast, negative formation energies correspond to the substitutional model (position c, see Fig.\ref{MoS2structure}), indicating that the substitution process is exothermic. In this configuration the Nb-S distances decrease by 0.05 \AA \- compared to the NbS$_2$ bulk, adapting to the initial parameters of the ``host'' MoS$_2$ lattice. The Mo-S bond lengths at neighboring sites remain unaffected.
In the mirror situation, when NbS$_2$ is doped with Mo, both the intercalation and substitution processes are energetically favorable.\footnote{Unlike the case of interstitial doping of MoS$_2$, the Mo tetrahedral interstitial in NbS$_2$ is metastable. The formation energy for the tetrahedral configuration is 1.87 eV higher than for the octahedral, thus in the following we will refer only to the case of octahedral interstitials.} Moreover, the range of formation energies reported in Table \ref{eformTab} suggests that low level substitutional and interstitial doping will be in competition.
The Mo-S bond distances for the substitutional Mo dopant increase by about 0.05 \AA \- compared to the MoS$_2$ bulk. However, Mo insertion between the NbS$_2$ planes is characterized by a decrease of  0.02 \AA \- in the Mo-S bond length.

\begin{figure*}[tbp]
  \includegraphics[bb=65 195 570 770, width=0.35\linewidth]{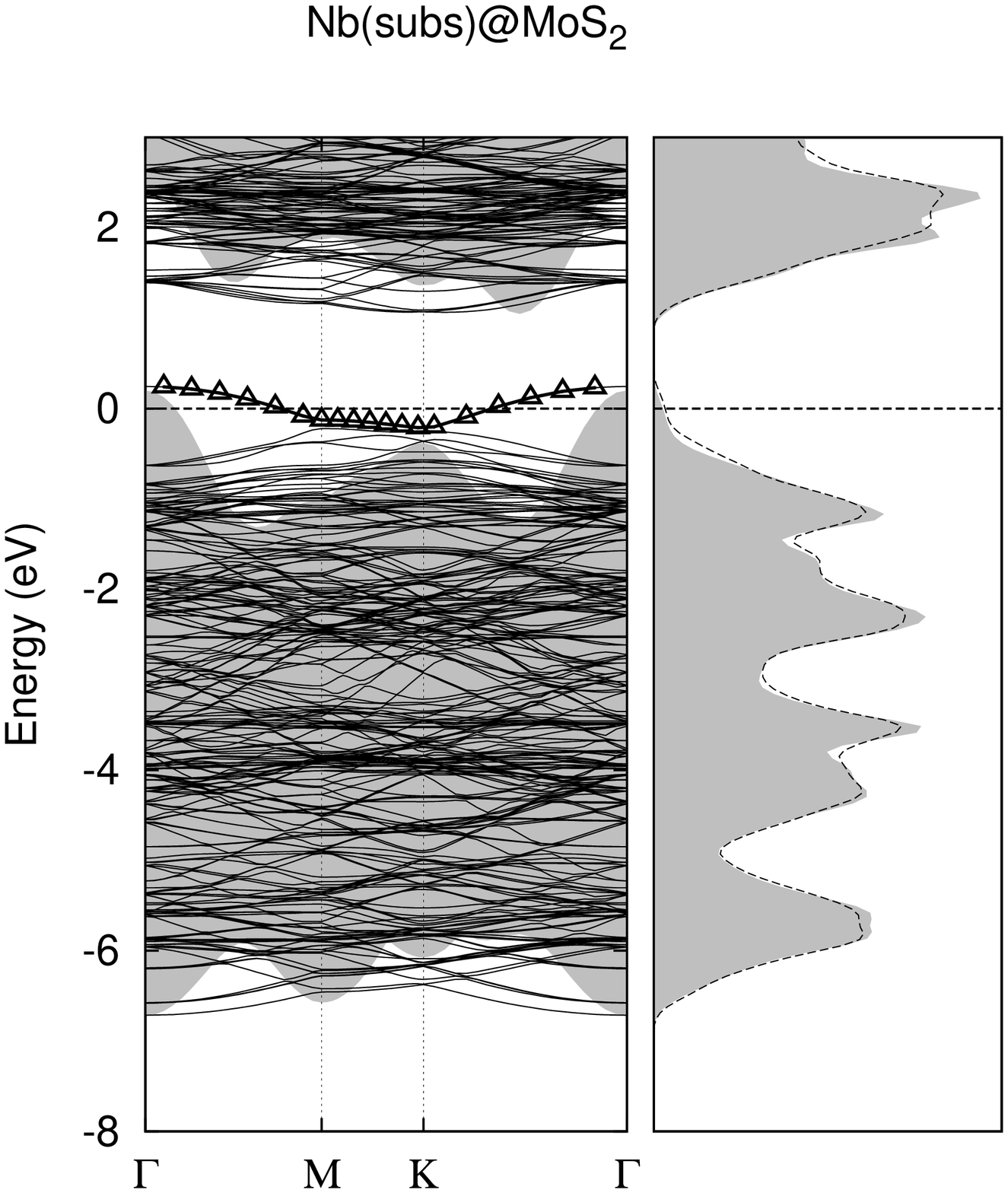}
  \includegraphics[bb=65 195 570 770,width=0.35\linewidth]{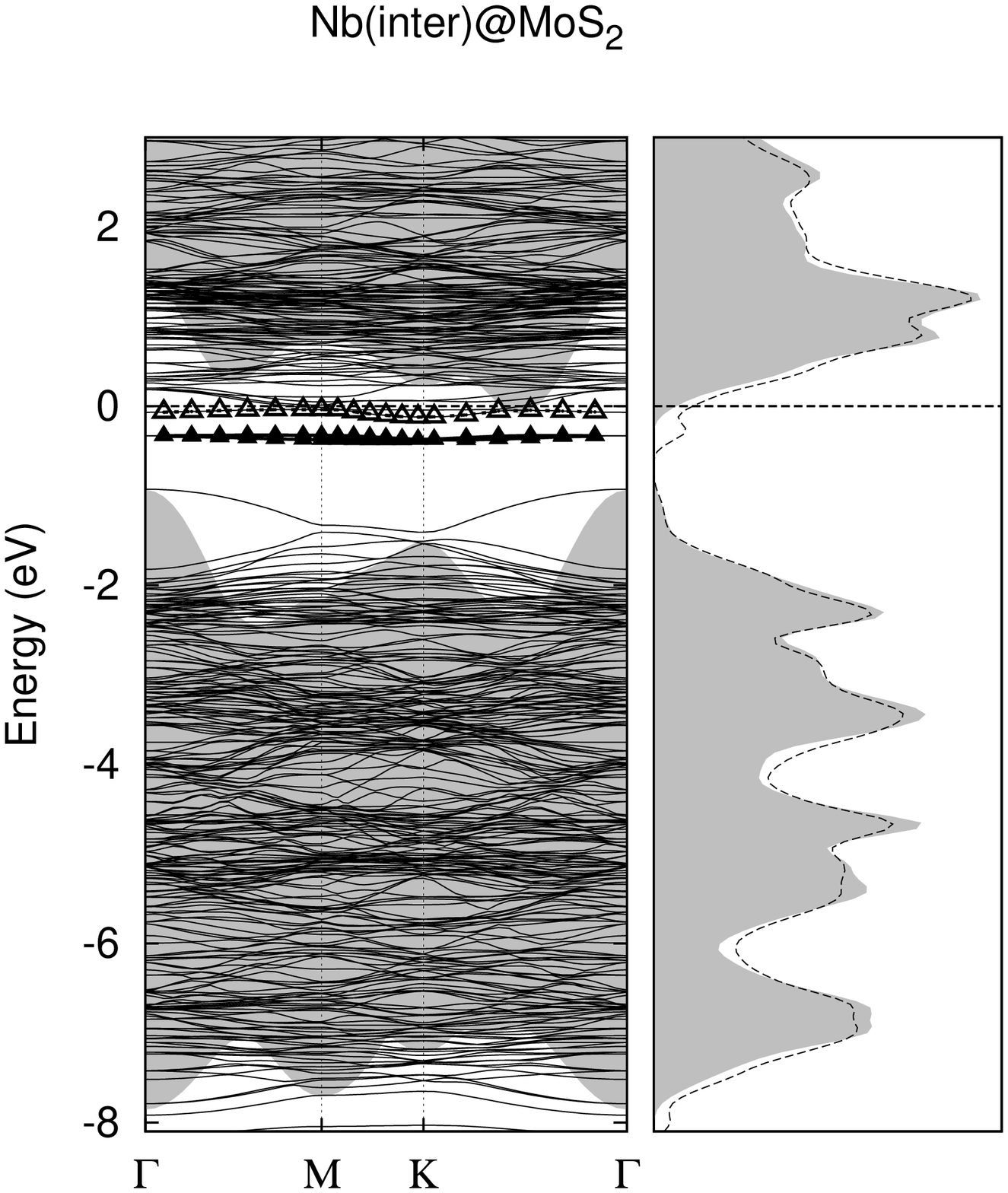}
  \includegraphics[bb=65 195 570 770,width=0.35\linewidth]{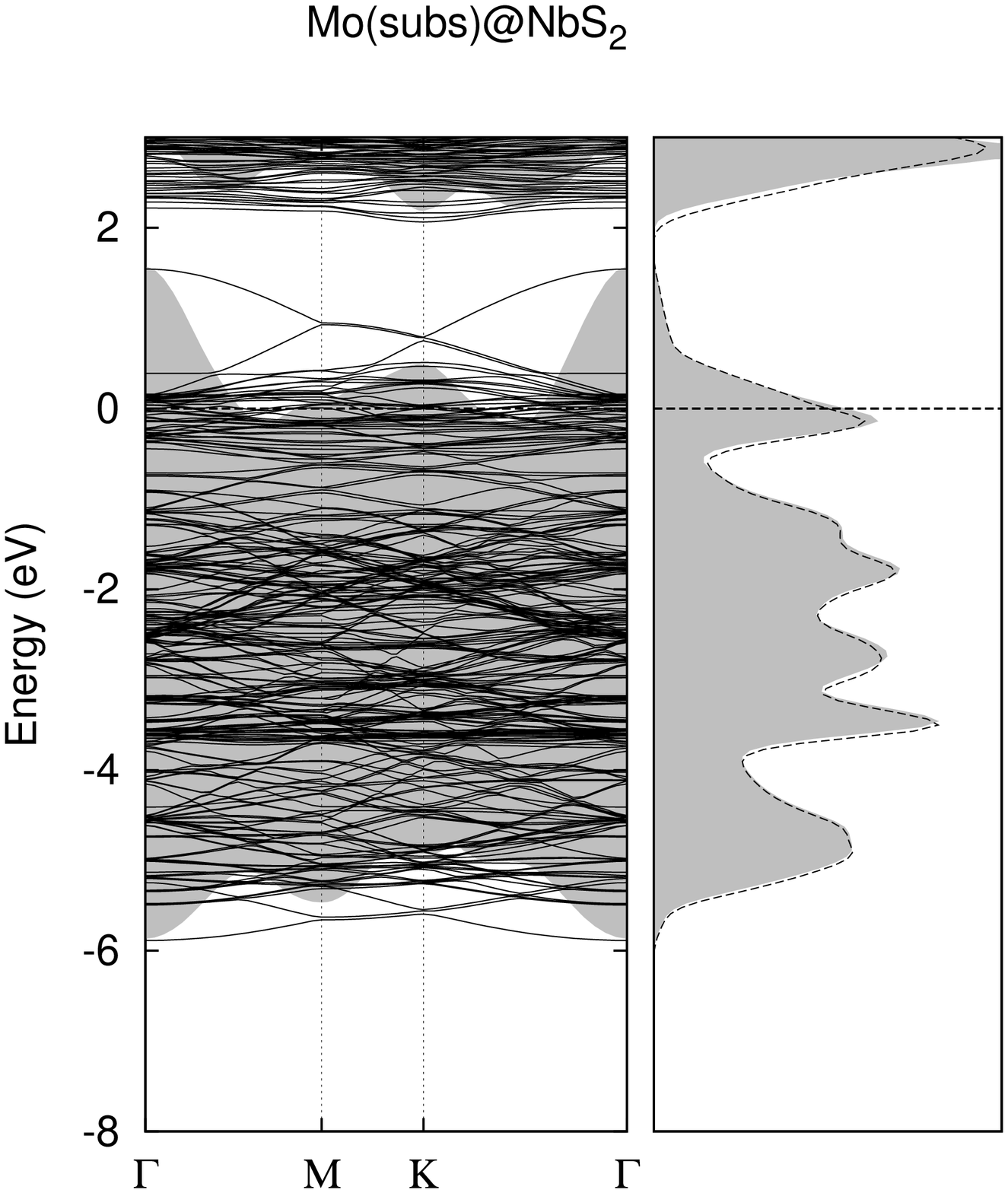}
  \includegraphics[bb=65 195 570 770,width=0.35\linewidth]{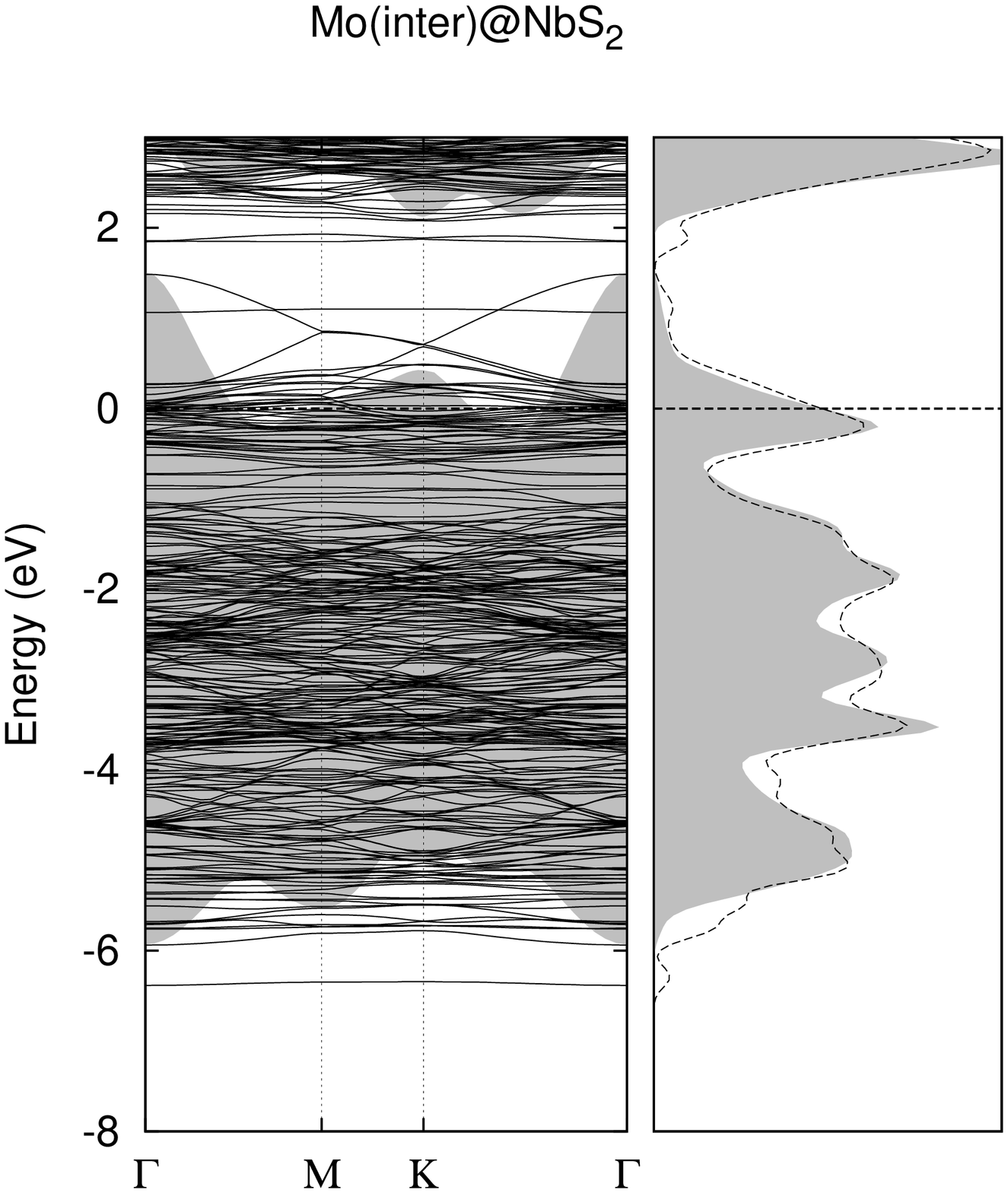}
  \caption{DFT-LDA electronic band structure and total density of states for substitutional and interstitial doping of MoS$_2$ (top) and NbS$_2$ (bottom) crystals. Filled areas corresponds to the pure bulk systems. Solid and open triangles indicate respectively the filled and half-filled electronic bands associated with the doping of MoS$_2$. The origin of the energies  in each graph is the Fermi level.}
  \label{subs1}
  \end{figure*}

\begin{figure}[b]
 \includegraphics[width=0.80\linewidth]{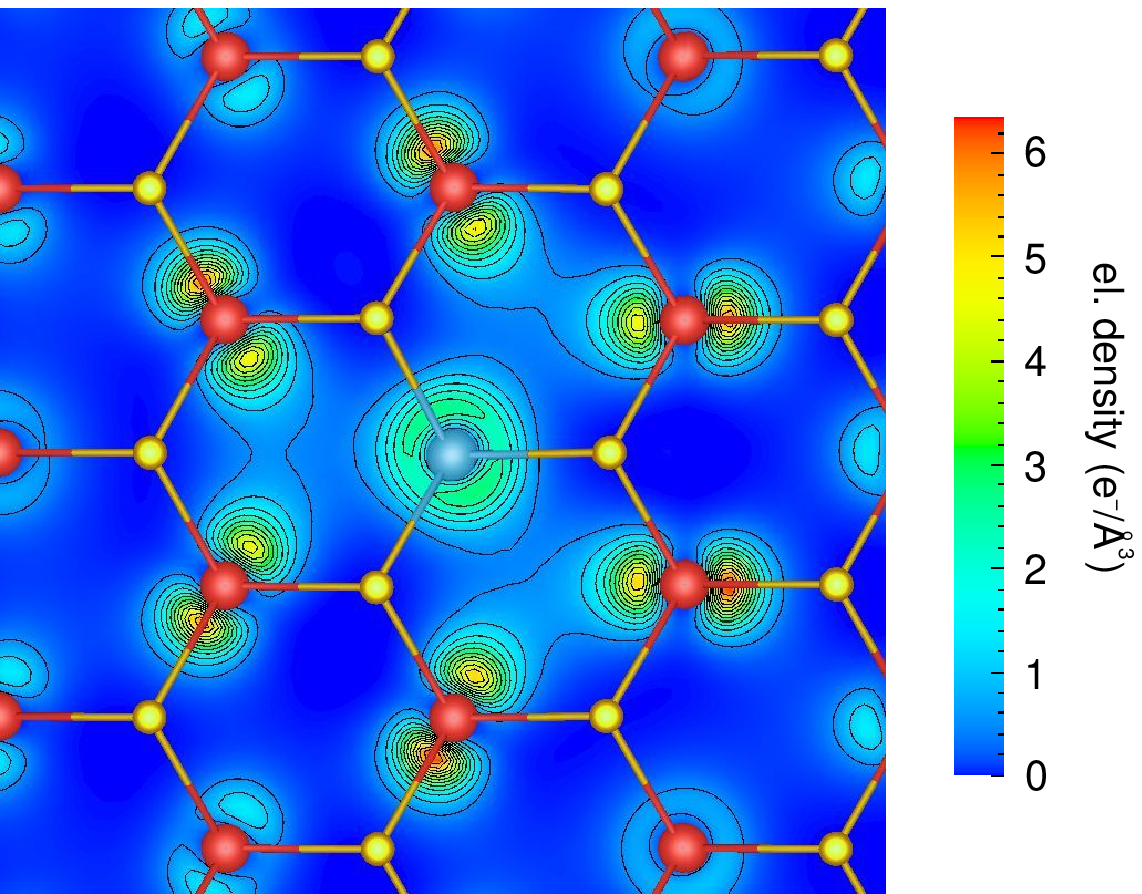}
  \caption{(Color online) Electron density distribution of the half-occupied electronic state  associated with the introduction of one Nb substitutional atom in the MoS$_2$ lattice. The Nb atom is blue, Mo atoms are red and S atoms are yellow.
}
  \label{final}
\end{figure}

In Fig. \ref{subs1} we present the band structure and density of states (DOS) for low level doping  of MoS$_2$ and NbS$_2$ crystals. The insertion of Nb dopant atoms within the MoS$_2$ preserves the main features of the DOS of the bulk system. Furthermore, the interstitial Nb atom adds two additional nearly degenerate fully occupied energy levels within the band gap and a half-occupied donor level at the bottom of the conduction band (see Fig. \ref{subs1}).
The substitution of one Mo atom by one Nb atom introduces an electron hole within the system and thus moves the  Fermi level down. The Nb atom acts as an acceptor impurity in the molybdenum disulphide. To get further insight into the electronic properties of Nb-substituted MoS$_2$, we examine the electron density distribution map for the the half-occupied electronic state associated solely with the doping, see Fig. \ref{final}. As one may observe, the impurity level is not centered exclusively on the Nb atom, but delocalized within the radius of the third molybdenum ring around the niobium, thus suggesting the metallic nature of the mixed system.

The substitutional doping of NbS$_2$ by Mo acts in a symmetric way to the previously treated case, introducing an additional electron instead of a hole into the system. Thus, the Fermi level is moved upward and the overall electronic structure of the system is conserved. Similar behavior is found for the interstitial doping with a slightly higher Fermi level shift.

\subsection{Higher dopant concentration: MoS$_2$-NbS$_2$ alloy}

In this section we treat the case of higher doping levels for the system MoS$_2$-NbS$_2$. Two extreme cases might be considered for the arrangement of dopant atoms within the original host matrix: dopants randomly distributed in the crystal forming a homogeneous alloy, or clusterized, forming local domains in the host crystal.

The tendency of dopants to clusterize can be investigated by considering the binding energy between two dopant atoms as a function of their separation. We discuss the situations when two dopants occupy close neighboring sites or, contrary to that, are situated at the farthest sites within the chosen supercell.
We define the binding energy E$_{\text{bind}}$ between dopants as:

\begin{equation}
E_{\text{bind}}=2\cdot E_{\text{1D}}-E_{\text{2D}}
\end{equation}

where E$_{\text{1D}}$ and E$_{\text{2D}}$ are respectively the formation energies of the systems with one and two dopants.

\begin{table*}[btp]
\caption{Formation energies E$_{\text{form}}$(eV) and binding energies E$_{\text{bind}}$ (eV) for  interstitial and substitutional doping of MoS$_2$ and NbS$_2$. Formation energies are calculated using the dopant chemical potentials derived from the metal disulphide.}
\begin{tabular*}{0.7\linewidth}{@{\extracolsep{\fill}}lrcr|lrr}
\hline
\hline
 System (interstitial) & E$_{\text{form}}$ & E$_{\text{bind}}$ & & System (substitutional) & E$_{\text{form}}$ & E$_{\text{bind}}$ \\
\hline
 NbMo$_{32}$S$_{64}$   &  8.08 & & & NbMo$_{31}$S$_{64}$  & -0.21  &     \\
 Nb$_2$Mo$_{32}$S$_{64}$  \footnotemark[1]&14.98 &  1.17 & & Nb$_2$Mo$_{30}$S$_{64}$\footnotemark[1]  &-0.44 &  0.01  \\
 Nb$_2$Mo$_{32}$S$_{64}$  \footnotemark[2] &16.07 &  0.08  & & Nb$_2$Mo$_{30}$S$_{64}$  \footnotemark[2] &-0.46 &   0.03  \\
 \hline
 MoNb$_{32}$S$_{64}$  &-3.24 &   & & MoNb$_{31}$S$_{64}$  &-5.11  &    \\
 Mo$_2$Nb$_{32}$S$_{64}$ \footnotemark[1]  & -0.91  &  -5.57& & Mo$_2$Nb$_{30}$S$_{64}$ \footnotemark[1] & -4.97 &  -5.25 \\
 Mo$_2$Nb$_{32}$S$_{64}$   \footnotemark[2]& -1.08  & -5.39  & & Mo$_2$Nb$_{30}$S$_{64}$  \footnotemark[2]& -4.91  &  -5.31 \\
\hline
\hline
\end{tabular*}
\footnotetext[1]{Two dopant atoms are separated by the shortest possible distance within the supercell}
\footnotetext[2]{Two dopant atoms are as far as possible from each other within the chosen supercell}
\label{bindEnTab}
\end{table*}

In Table \ref{bindEnTab} we report formation and binding energies for the mentioned arrangements of dopants within the MoS$_2$ and NbS$_2$ lattices.
For interstitial Nb atoms in the  MoS$_2$ crystal we obtain a high binding energy, which at two neighboring octahedral sites rises to 1.17 eV. This value suggests a strong driving force which promotes the clustering of Nb interstitials. In contrast, Nb substitutional atoms weakly interact, with binding energies of the order of a hundredth of an eV. Considering that the configurational entropy is higher for a random distribution of dopant atoms, we thus conclude that the mixed system will tend to form a homogeneous alloy.  A similar tendency for random ordering is obtained for both interstitial and substitutional Mo in NbS$_2$, where the binding energies between two doping atoms drop to about $\sim -5$ eV.

The experimental work of Hotje et al.\cite{Binnewies-05} has shown that the MoS$_2$-NbS$_2$ alloy can be synthesized  without any miscibility gap across the concentration range going from pure MoS$_2$ to pure NbS$_2$.
We have thus extended our investigation to higher doping levels up to a maximum of 25\%. The percentage of doping is calculated considering all the metal atoms within the two adjacent metal disulphide slabs, constituting the supercell.

\begin{figure}
 \includegraphics[bb=85 130 487 728, angle=-90,width=\linewidth]{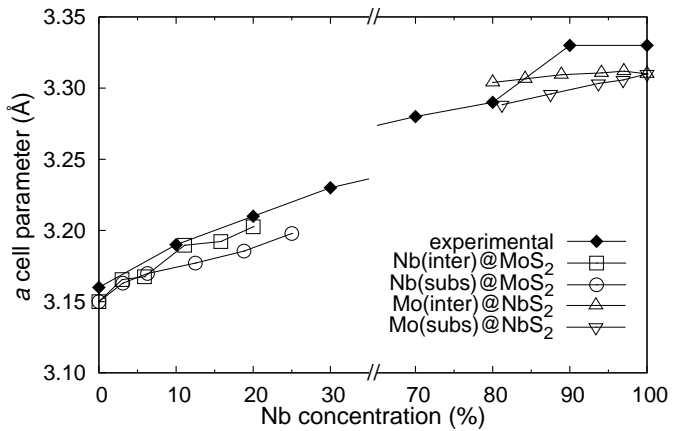}
  \caption{In-plane cell parameter for the mixed Mo$_{1-x}$Nb$_x$S$_2$ systems as a function of Nb concentration inside the cell. Solid diamonds correspond to reported experimental data. \cite{Binnewies-05}}
  \label{Graph2}
\end{figure}

In section \ref{structModel} we have seen that the in-plane lattice parameters of MoS$_2$ and NbS$_2$ differ by about 4.8 \%. Then, in section \ref{secLowcon} we saw that the incorporation of dopants causes structural perturbations related to different metal-sulphur bond lengths. For high doping concentrations these frustrations will induce a change in the crystallographic parameters of the host crystal which depends on the stoichiometries of the mixed systems.
The data, obtained by optimizing the atomic positions simultaneously with the cell parameters, are presented in Fig. \ref{Graph2} and compared to the X-ray diffraction data available from the work of Hotje et al.\cite{Binnewies-05} The graph shows a good agreement between our theoretical results and the experimental data with a deviation of around 1\% across the range of doping levels considered.
\footnote{In Ref. \cite {Binnewies-05} the cell parameter values reported refer to the 3R-NbS$_2$ phase, and partially to 2H and 3R MoS$_2$. Since the difference in the \emph{a} cell parameter  between the 2H and the 3R phases is as low as 0.13 and 0.60 \% for MoS$_2$ and NbS$_2$ respectively, in Fig. \ref{Graph2} we compare the experimental values with our results derived only for the 2H phases.}

\begin{figure}[btp]
 \includegraphics[bb=85 130 487 728, angle=-90,width=\linewidth]{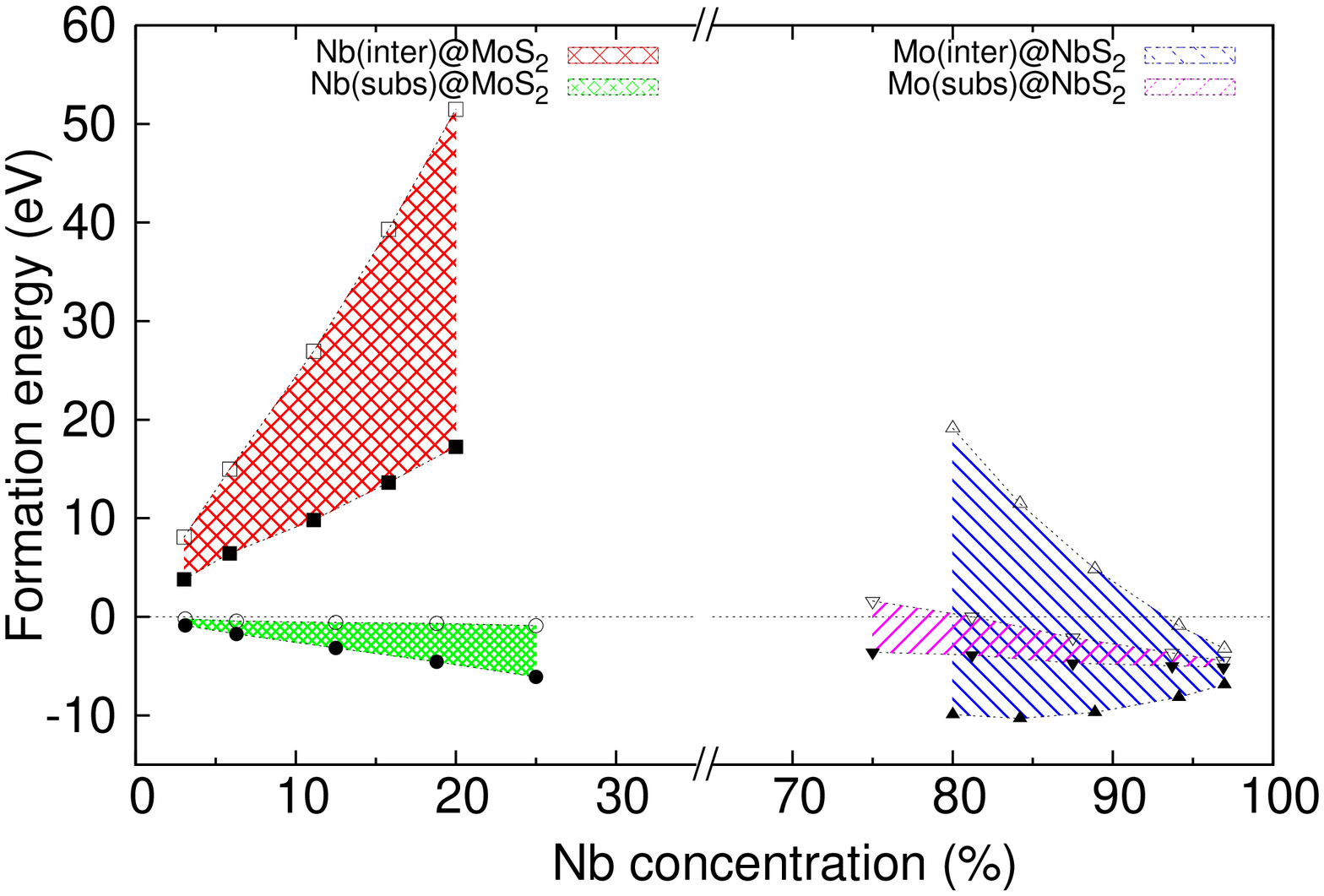}
  \caption{(Color online) Formation energies for the mixed Mo$_{1-x}$Nb$_x$S$_2$ systems as a function of the dopant chemical potential and dopant concentration within the cell. Open symbols denote the dopant chemical potential derived from pure metal disulphide, solid symbols stand for potential obtained from metal bulk phases.}
  \label{Graph3}
\end{figure}

Formation energies for the mixed MoS$_2$-NbS$_2$ systems as a function of interstitial and substitutional dopant concentration are presented in Fig. \ref{Graph3}.
The range of values represented by the shaded areas and obtained from Eq. \ref{EformInter}, is defined by the two dopant chemical potentials of the pure metal disulphide and metal bulk phases. The observed trends in E$_{\text{form}}$ for low level Nb doping of MoS$_2$ are preserved as well at higher dopant concentrations showing a preferential substitutional doping. These results are confirmed by recent EXAFS experiments.\cite{Catal}

Now we turn to the symmetric situation of Mo doping of NbS$_2$. As we observed before, low level doping by Mo substitution or intercalation within the NbS$_2$ are exothermic (and competing) processes. The enhancement of dopant content leads to the overlapping of the regions corresponding to the formation energies  for the interstitial and substitutional doping, see Fig. \ref{Graph3}. Hence, depending upon the particular synthesis conditions, one may have a combination of substituted and intercalated molybdenum atoms within the NbS$_2$ matrix.

The effect of low level doping of MoS$_2$ and NbS$_2$ crystals on their electronic properties have been discussed in detail in Sec. \ref{secLowcon}.
Here, we mention that near the Fermi level the electronic structure is locally sensitive not only to the type but also to the degree of doping. However, the main tendencies described above will be preserved and enhanced by higher dopant concentrations. Thus, further Nb substitution in  MoS$_2$ will decrease the total electron concentration of the system and cause a progressive downward shift of the Fermi level. However, the substitutional doping with Mo of NbS$_2$, will increase the total electron concentration and lead to gradual filling of the $d_{z^2}$ Nb band.

Hence, over the whole range of considered concentration, the electronic behavior of the stable mixed systems will be metallic. We note that our conclusions agree well with the experimental data. According to electrical resistivity  measurements, MoS$_2$ with 5 \% of substitutional Nb atoms has an electrical resistivity level similar to that of graphite.\cite{Kalikhman-82} Likewise, the mixed Nb$_x$Mo$_{1-x}$S$_2$ nanoparticles are reported to be metallic.\cite{Deepack-07}

\section{Conclusions}

We have presented a systematic \textit{ab initio} study on the stability, structural and electronic properties of mixed molybdenum niobium disulphides.
Bilateral doping effects have been investigated up to a dopant concentration of 25 \%.

Focusing on the specific arrangement of dopant atoms, we observe that over the whole range of considered concentrations, substitutional doping with Nb of MoS$_2$ will predominate. In addition, our calculations for Mo doping of NbS$_2$ show that depending on the specific synthesis conditions, both interstitial and substitutional Mo arrangements can co-exist. The incorporation of dopant atoms causes structural perturbations and changes in the crystallographic parameters of the host crystal, which dependents on the stoichiometries of the mixed systems.

The difference in the number of valence electrons between Mo and Nb atoms means that Nb substitutional doping of MoS$_2$ introduces electron holes into the system. Mo substitutional doping of NbS$_2$ adds electrons to the system, leading to an upward shift of the Fermi level. According to our results, these mixed disulphides have a metallic behavior throughout the range of stoichiometry considered. This result is in agreement with experimental data available for low-level Nb doping of MoS$_2$.

We should note that the semiconductor to metal transition due to Nb doping might not induce any degradation in the mechanical properties of the MoS$_2$. Indeed, the dopant atoms are located within the planes and no inter plane bonding bridges are formed. Thus, mixed NbS$_2$-MoS$_2$ might present interesting capabilities for new tribological applications.

The conclusions of the present work can be generalized to other mixed \textit{d} metal chalcogenides presenting similar structural and electronic properties.

\section{Acknowledgements}

This work was financed  by the FOREMOST project of the European Union 6-th Framework Program under contract NMP3-CT-2005-515840. V.V.I. thanks the Foundation of the President of Russian Federation (Grant-502.2008.3) for financial support. The authors would like to acknowledge G. Seifert for interesting discussions and M. Walls for reading the manuscript.

\end{document}